\documentclass[twocolumn,showpacs,preprintnumbers,amsmath,amssymb,prb]{revtex4}

\usepackage{graphicx}
\usepackage{dcolumn}
\usepackage{bm}
\usepackage{tabularx}
\usepackage{amsmath}

\begin{document}

\title{Low-temperature phonon thermal conductivity of cuprate single crystals}

\author{S. Y. Li,$^1$ J.-B. Bonnemaison,$^1$ A. Payeur,$^1$ P. Fournier,$^{1,2}$ C. H. Wang,$^3$ X. H. Chen,$^3$ and Louis Taillefer$^{1,2,*}$}

\affiliation{$^1$D{\'e}partement de physique and RQMP,
Universit{\'e} de Sherbrooke, Sherbrooke, Canada\\
$^2$Canadian Institute for Advanced Research, Toronto, Canada\\
$^3$Hefei National Laboratory for Physical Science at Microscale and
Department of Physics, University of Science and Technology of
China, Hefei, Anhui 230026, P. R. China}

\date{\today}

\begin{abstract}
The effect of sample size and surface roughness on the phonon
thermal conductivity $\kappa_p$ of Nd$_2$CuO$_4$ single crystals was
studied down to 50 mK. At 0.5 K, $\kappa_p$ is proportional to
$\sqrt{A}$, where $A$ is the cross-sectional area of the sample.
This demonstrates that $\kappa_p$ is dominated by boundary
scattering below 0.5 K or so. However, the expected $T^3$ dependence
of $\kappa_p$ is not observed down to 50 mK. Upon roughing the
surfaces, the $T^3$ dependence is restored, showing that departures
from $T^3$ are due to specular reflection of phonons off the
mirror-like sample surfaces. We propose an empirical power law fit,
to $\kappa_p \sim T^{\alpha}$ (where $\alpha < 3$) in cuprate single
crystals. Using this method, we show that recent thermal
conductivity studies of Zn doping in YBa$_2$Cu$_3$O$_y$ re-affirm
the universal heat conductivity of $d$-wave quasiparticles at $T \to
0$.

\end{abstract}


\maketitle

To understand the pairing mechanism in a superconductor, it is
essential to know the symmetry of the order parameter. In this
context, measurements of low-temperature thermal conductivity
$\kappa$, which probes the low-energy quasiparticle excitations, has
emerged as a powerful probe of the order parameter in
superconductors. For conventional $s$-wave superconductors with a
fully gapped excitation spectrum, the linear-temperature electronic
contribution to thermal conductivity is zero at $T \to 0$, i.e. the
residual linear term $\kappa_0/T = 0$. This can be seen in the
single-gap $s$-wave superconductor Nb, \cite{Lowell} multi-band
$s$-wave superconductors MgB$_2$ (Ref. 2) and NbSe$_2$,
\cite{Boaknin} and two newly discovered superconductors C$_6$Yb
(Ref. 4) and Cu$_x$TiSe$_2$. \cite{Li}

For unconventional superconductors with nodes in the gap, the nodal
quasiparticles will contribute a finite $\kappa_0/T$. For example,
$\kappa_0/T$ = 1.41 mW K$^{-2}$ cm$^{-1}$ in the overdoped cuprate
Tl2201, a $d$-wave superconductor with $T_c$ = 15 K, \cite{Proust}
and $\kappa_0/T$ = 17 mW K$^{-2}$ cm$^{-1}$ for the ruthenate
Sr$_2$RuO$_4$, a $p$-wave superconductor with $T_c$ = 1.5 K.
\cite{Suzuki} The fact that $\kappa_0/T$ is universal,
\cite{Taillefer} in the sense that it is independent of scattering
rate, allows for a measurement of the gap via \cite{Hawthorn1}
\begin{equation}
\frac{\kappa_0}{T} \simeq \frac{{k_B}^2}{6}
\frac{n}{c}\kappa_F\frac{v_F}{\Delta_0}.
\end{equation}
This has been used to map out the gap as a function of doping in
YBa$_2$Cu$_3$O$_y$ \cite{Sutherland1} and Tl2201. \cite{Hawthorn1}

The measured thermal conductivity is the sum of two contributions,
respectively from electrons and phonons, so that $\kappa = \kappa_e
+ \kappa_p$, where $\kappa_e/T$ is a constant as $T \to 0$.
Therefore the key issue for these low-temperature thermal
conductivity studies is how to extrapolate $\kappa/T$ to $T = 0$,
i.e. to extract the residual linear term $\kappa_0/T$. This requires
a good understanding of the phonon conductivity $\kappa_p(T)$. In
the regime $T \to 0$, the phonon mean free path becomes limited only
by the physical dimensions of the sample. At the surface of a
crystal, the phonon may either be absorbed and reemitted with an
energy distribution given by the local temperature (diffuse
scattering) or it may be reflected elastically (specular
reflection). In the case of diffuse scattering, the phonon is
reradiated in a random direction resulting in a temperature
independent phonon mean free path $l_0$, given by the
cross-sectional area $A$ of the sample: $l_0 = 2\sqrt{A/\pi}$. From
simple kinetic theory, the conductivity of phonons is given by
\cite{Casimir}
\begin{gather}
\kappa_p = \frac{1}{3} \beta<v_p> l_0 T^3,
\end{gather}
where $\beta$ is the coefficient of phonon specific heat, and
$<v_p>$ is a suitable average of the acoustic sound velocities. The
electronic linear term is then naturally extracted by plotting
thermal conductivity data as $\kappa/T$ vs $T^2$ and interpreting
the intercept as the residual linear term at $T = 0$, and the slope
as the phonon contribution governed by Eq. (2).

However, as the temperature of a crystal is reduced and the average
phonon wavelength increases, a surface of a given roughness appears
smoother, which may increase the occurrence of specular reflection
and result in a mean free path which varies as some power of
temperature, so that $\kappa_p \propto T^\gamma$. We would thus
expect a deviation from the diffuse scattering limit of $T^3$
temperature dependence for samples with sufficiently smooth
surfaces. Such an effect has been previously observed in many
studies of low-temperature phonon heat transport in high quality
crystals, such as Al$_2$O$_3$, \cite{Pohl} Si, \cite{Hurst} KCl and
KBr, \cite{Seward} LiF, \cite{Thatcher} and diamond. \cite{Berman}

For the high-$T_c$ cuprate superconductors, the extrapolation of
$\kappa_0/T$ has been a controversial issue, particularly in the
underdoped regime where $\kappa_0/T$ is small. Some authors fit
their data to $\kappa/T = a + bT^2$ below about 120 mK, assuming
that boundary scattering of phonons only occurs below 120 mK and
there is no specular reflections. \cite{Sun1,Sun2} Others consider
that sample size and specular reflections do affect $\kappa_p$, and
fit their data to $\kappa/T = a + bT^\alpha$ below about 0.5 K.
\cite{Sutherland1,Sutherland2,Nicolas} Disagreement on extrapolation
procedure has fueled a debate on whether YBa$_2$Cu$_3$O$_y$ is a
thermal metal below the doping level of $p =$ 0.05,
\cite{Sutherland2,Nicolas,Sun1} and possible breakdown of the
universal thermal conductivity in YBa$_2$Cu$_3$O$_7$ and
YBa$_2$Cu$_3$O$_{6.5}$. \cite{Sun2}

In this paper, we investigate the low-temperature phonon thermal
conductivity of cuprate single crystals by measuring the insulating
(undoped) parent compound Nd$_2$CuO$_4$, for which $\kappa_0/T = 0$.
First, by reducing the sample width $w$, we show that $\kappa_p$ is
proportional to $\sqrt{wt}$ at 0.5 K, with $t$ the unchanged sample
thickness. This clearly demonstrates that $\kappa_p$ is dominated by
boundary scattering below 0.5 K. Secondly, we show a sample with
rough surfaces has a temperature dependence much closer to $T^3$
than the one with smooth surfaces, which is apparently caused by the
reducing of specular reflections at the rough surfaces. Finally,
based on this understanding of $\kappa_p$, we discuss how to extract
$\kappa_0/T$ in cuprate single crystals at finite doping (where
$\kappa_0/T > 0$).

Single crystals of Nd$_2$CuO$_4$ were grown by a standard flux
method. The as-grown samples are plate-like with very smooth
surfaces in the $ab$-plane. In-plane thermal conductivity $\kappa$
was measured in a dilution refrigerator down to 50 mK using a
standard one heater-two thermometer steady-state technique. Note
that in zero field, magnons are gapped and the thermal conductivity
of insulating Nd$_2$CuO$_4$ only comes from phonons: $\kappa$ =
$\kappa_p$. \cite{Li2}

\begin{figure}
\includegraphics[clip,width=8.5cm]{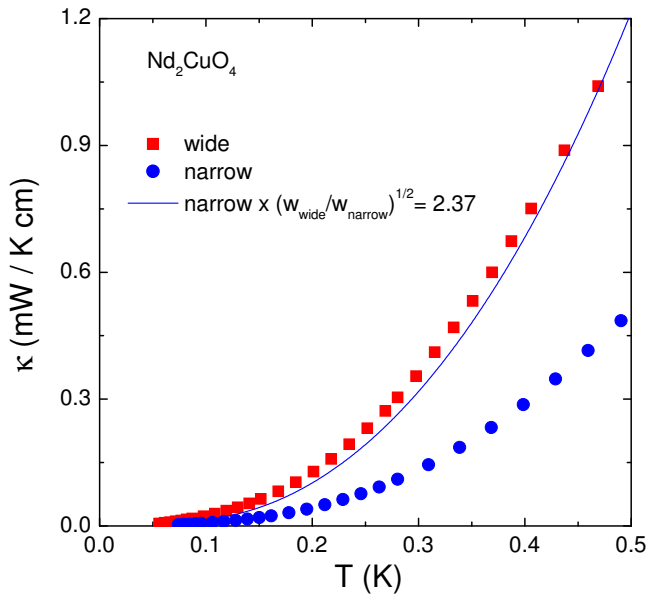}
\caption{(color) In-plane thermal conductivity $\kappa$ of the
undoped insulating cuprate Nd$_2$CuO$_4$ below 0.5 K, for an
as-grown single crystal before (``wide"; red squares) and after
(``narrow"; blue circles) it was cut along the length. By cutting,
the width $w$ was reduced by a factor of 5.6. At $T$ = 0.5 K,
$\kappa$ is reduced in direct proportion to the reduction in
$\sqrt{w}$, as indicated by the solid line.}
\end{figure}

{\it Boundary scattering.} ---  To study the boundary scattering of
phonons, an as-grown Nd$_2$CuO$_4$ single crystal with dimensions
1.5 $\times$ 0.90 $\times$ 0.086 mm$^3$ (length $\times$ width
$\times$ thickness) was measured first. Then it was cut along the
length so that its width $w$ was reduced (by a factor of 5.6) to
0.16 mm, and measured again. Fig. 1 shows $\kappa$ vs $T$ for the
wide (before cutting) and narrow (after cutting) samples below 0.5
K. It is clear that the narrow sample has a much smaller $\kappa$
than the wide sample in this temperature range. Actually, at $T$ =
0.5 K, $\kappa$ of the narrow sample is precisely $\sqrt{5.6} =
2.37$ times smaller than the wide sample. This means $\kappa$ is
reduced in direct proportion to the reduction in $\sqrt{A} =
\sqrt{wt}$, where $t$ is the unchanged sample thickness. This is
strong evidence that $\kappa$ is limited by sample size, and
boundary scattering is unambiguously the dominant mechanism for
phonon conductivity below 0.5 K.

\begin{figure}
\includegraphics[clip,width=8.5cm]{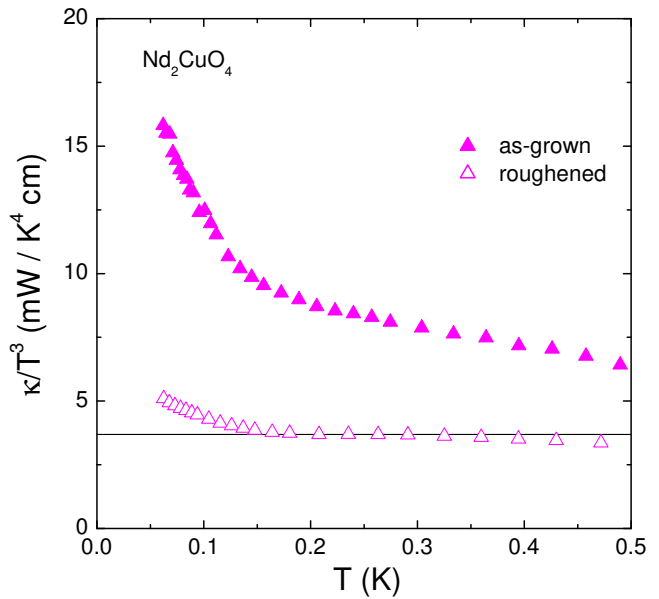}
\caption{(color) In-plane thermal conductivity $\kappa$ of an
as-grown single crystal of Nd$_2$CuO$_4$ before (``as-grown"; full
triangles) and after (``roughened"; empty triangles) its smooth
mirror-like surfaces were roughened by sanding. As indicated by the
horizontal line, roughening allows one to recover the diffusive
regime at low temperature, characterized by a constant $\kappa/T^3$,
at least down to 150 mK.}
\end{figure}

{\it Specular reflection}. --- Having demonstrated that phonons are
scattered by sample surfaces below 0.5 K, we proceed to study the
effect of surface roughness on the temperature dependence of
$\kappa$. An as-grown Nd$_2$CuO$_4$ single crystal with dimensions
1.0 $\times$ 0.43 $\times$ 0.086 mm$^3$ was measured first.
Afterwards, both its mirror-like $ab$-plane surfaces were roughened
by sanding, then it was re-measured. During sanding, the thickness
of the sample was reduced. The geometric factor for the roughened
sample was set to be such that $\kappa$(rough) = $\kappa$(smooth) at
high temperature (e.g. 50 K). In Fig. 2, the data of the smooth and
roughened samples are plotted as $\kappa/T^3$ vs $T$, to reveal the
deviation from the $T^3$ dependence expected if the phonon mean free
path $l_p \propto \kappa/T^3$ were constant, independent of $T$.
$l_p$ is clearly not constant for the smooth sample, while for the
roughened sample, roughening has made $l_p$ much more constant, at
least down to 150 mK. This is unambiguous proof that specular
reflection is important in these crystals. Below 150 mK, the phonon
wavelength becomes long enough to average over the roughness and
produce some specular reflections.

{\it Extracting $\kappa_0/T$}. --- From Figures 1 and 2, it is clear
that for typical cuprate single crystals with smooth surfaces,
phonon thermal conductivity will reach the boundary scattering
regime below 0.5 K and $\kappa$ is expected to deviate from the
standard $T^3$ dependence due to specular reflections. Considering
together previous studies of other conventional insulators,
\cite{Pohl,Hurst,Seward,Thatcher,Berman} a reasonable way to extract
$\kappa_0/T$ of cuprate single crystals is to fit the thermal
conductivity data to $\kappa/T = a + bT^\alpha$. In Fig. 3, we
reproduce two sets of published data: one on undoped Nd$_2$CuO$_4$
(from Ref. 21), the other on fully doped YBa$_2$Cu$_3$O$_7$ (From
Ref. 18). For Nd$_2$CuO$_4$, fitting $\kappa/T$ to $a + bT^\alpha$
below 175 mK gives $\kappa_0/T$ = -0.004 $\pm$ 0.006 mW K$^{-2}$
cm$^{-1}$ and $\alpha$ = 1.29 $\pm$ 0.04. In contrast, a fit to
$\kappa/T = a + bT^2$ below 120 mK gives $\kappa_0/T$ = 0.054 $\pm$
0.003 mW K$^{-2}$ cm$^{-1}$. For this insulator, the negligible
$\kappa_0/T$ obtained from the first fit is physically more
reasonable than the finite $\kappa_0/T$ from the second fit. This
shows that a textbook fit overestimates $\kappa_0/T$.

\begin{figure}
\includegraphics[clip,width=8.5cm]{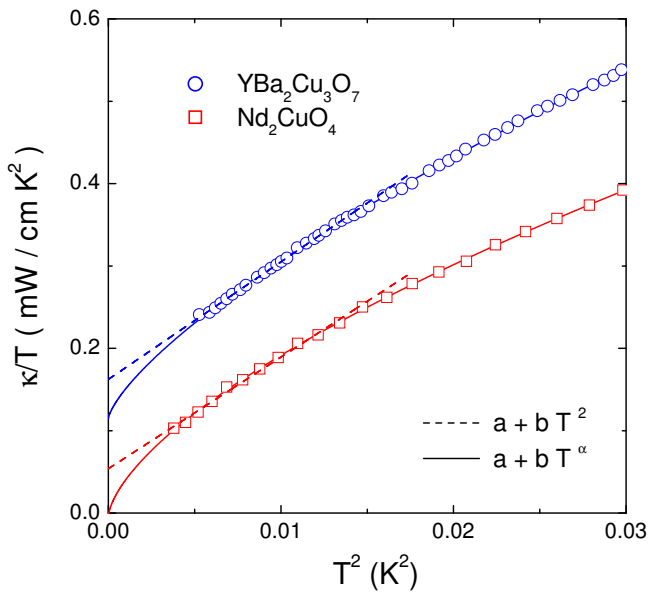}
\caption{(color) In-plane thermal conductivity $\kappa$ of undoped
Nd$_2$CuO$_4$ (data from Ref. 21) and slightly overdoped
YBa$_2$Cu$_3$O$_7$ (data from Ref. 18) single crystals, plotted as
$\kappa/T$ vs $T^2$ below 175 mK. The dashed and solid lines
represent two extrapolation procedures: 1) $\kappa/T = a + bT^2$
below 120 mK and 2) $\kappa/T = a + bT^\alpha$ below 175 mK. The
former yields a spurious residual linear term ($\kappa_0/T > 0$) in
the insulator Nd$_2$CuO$_4$, and thus overestimates $\kappa_0/T$ in
the superconductor YBa$_2$Cu$_3$O$_7$.}
\end{figure}

\begin{figure}
\includegraphics[clip,width=7cm]{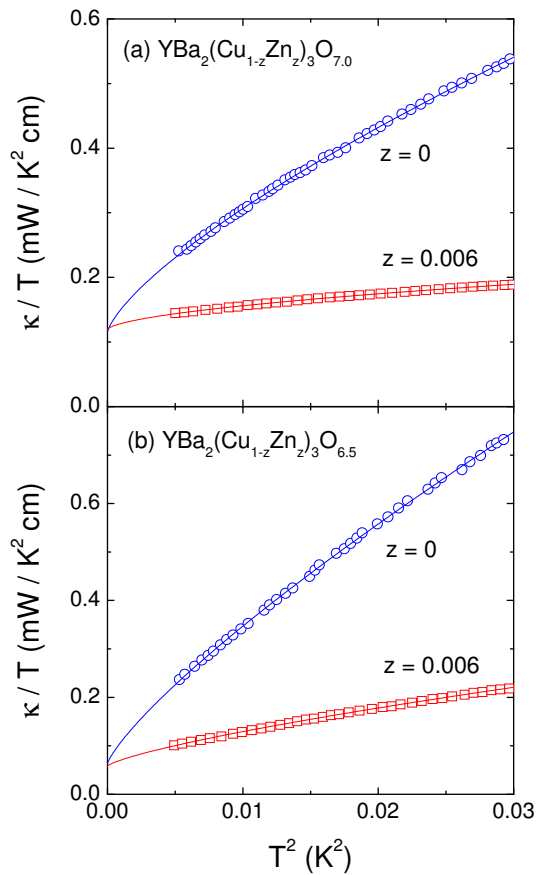}
\caption{(color) Thermal conductivity of pure and Zn-substituted
YBa$_2$Cu$_3$O$_y$ with (a) $y$ = 7.0 and (b) $y$ = 6.5 (data from
Ref. 18). The solid lines are fits to $\kappa/T = a + bT^\alpha$
below 175 mK, used to extract $\kappa_0/T$, seen to be universal in
both cases (i.e. unchanged by Zn substitution).}
\end{figure}

Applying the same two fits to YBa$_2$Cu$_3$O$_7$ (dashed and solid
lines in Fig. 3) gives $\kappa_0/T$ = 0.162 $\pm$ 0.001 and 0.115
$\pm$ 0.004 mW K$^{-2}$ cm$^{-1}$, respectively. Based on our
findings of Nd$_2$CuO$_4$ (Fig. 3), the lower estimate ($\kappa_0/T$
= 0.115 mW K$^{-2}$ cm$^{-1}$) is expected to be closer to the true
value for YBa$_2$Cu$_3$O$_7$.

Let us now turn to the controversial issue of whether thermal
conductivity is universal in YBa$_2$Cu$_3$O$_y$. In a $d$-wave
superconductor, $\kappa_0/T$ is due to nodal quasiparticles, and
standard theory shows $\kappa_0/T$ to be ``universal",
\cite{Graf,Durst} i.e. independent of impurity concentration, in the
limit of weak scattering ($\Gamma \ll \Delta_0$). Upon increasing
the scattering rate $\Gamma$ so that it becomes a significant
fraction of the gap maximum $\Delta_0$, $\kappa_0/T$ is expected to
increase slightly \cite{Graf,SunandMaki} (assuming the normal state
is a metal). Experimentally, it was first observed in
optimally-doped cuprates: YBa$_2$Cu$_3$O$_y$ ($y$ = 6.9) as a
function of Zn doping, \cite{Taillefer} and
Bi$_2$Sr$_2$CaCu$_2$O$_{8-x}$ as a function of radiation damage.
\cite{Nakamae} It was later verified in the layered nodal
superconductor Sr$_2$RuO$_4$. \cite{Suzuki}

Repeating the original experiment, Sun {\it et al.} \cite{Sun2}
reported $\kappa$ data for a pure and a Zn-doped crystal of
YBa$_2$Cu$_3$O$_y$ with $y = 7.0$, reproduced in Fig. 4a. The main
effect of Zn doping is to cause a dramatic suppression of the {\it
slope} of $\kappa / T$ as a function of $T^{2}$. The authors
extrapolate the data using a fit to $\kappa / T = a + bT^2$,
restricted to a very small interval (70 to 120 mK). The value of
$\kappa_0 / T$ thus extrapolated turns out to be 20\% lower in the
sample with 0.6\% Zn. Based on this slight difference, Sun {\it et
al.} go on to claim a breakdown of universal transport in
YBa$_2$Cu$_3$O$_y$.

However, the determination of  $\kappa_0 / T$ depends on the
extrapolation procedure, as seen in Fig. 3. Using a fit to $\kappa/T
= a + bT^\alpha$ below 170 mK for pure and 0.6\% Zn doped
YBa$_2$Cu$_3$O$_7$ (solid lines in Fig. 4a) yields $\kappa_0 / T$  =
0.115 and 0.119 mW K$^{-2}$ cm$^{-1}$, respectively. This means that
within error bars their data confirms the validity of the standard
theory of universal transport in this archetypal cuprate. In Fig.
4b, a similar result is found in underdoped YBa$_2$Cu$_3$O$_{6.5}$,
where $\kappa_0/T$ = 0.063 and 0.058 mW K$^{-2}$ cm$^{-1}$ for pure
and 0.6\% Zn doped.

Any variation in $\kappa_0/T$ with Zn doping should be put in proper
context: with 0.6\% Zn doping, the inelastic scattering rate was
estimated to increase by roughly a factor 10 in
YBa$_2$Cu$_3$O$_{6.9}$. \cite{Taillefer} Hence, the normal-state
$\kappa / T$ in the zero-temperature limit should decrease by a
factor of about 10. Therefore, the change in quasiparticle
conductivity measured in the superconducting state, if any, is seen
to be at most a few percent of the change expected in the normal
state. This is precisely what is meant by universal transport.

In contrast to YBa$_2$Cu$_3$O$_y$, the breakdown of standard theory
for La$_{2-x}$Sr$_2$CuO$_4$ (LSCO) \cite{Sun2} appears to be fairly
unambiguous (a rigid shift in $\kappa / T$ caused by a small change
in scattering rate) and consistent with a prior report of breakdown
in underdoped LSCO. \cite{Sutherland1} This is presumably a
consequence of the fact that suppressing the superconducting state
in LSCO leads to an insulating (and magnetic) state, rather than the
metallic state assumed by standard theory. \cite{Hawthorn,Sun3}

In summary, by studying the effect of sample size and surface
roughness on the phonon thermal conductivity of Nd$_2$CuO$_4$ single
crystals, we show that the phonon heat conduction in cuprates is
dominated by boundary scattering below 0.5 K. In as-grown (or
polished) single crystals, specular reflection alters the $T$
dependence away from the expected $T^3$ dependence. As a result, in
no range of temperature down to $T$ = 0 is a $T^3$ fit to the phonon
part of $\kappa$ appropriate. A better, but by no means exact, fit
is obtained by allowing the power to adjust away from 3, towards 2,
as found in conventional insulators. In the absence of a good
theoretical treatment of specular reflection, this is probably the
best one can do.

We thank K. Behnia for useful discussions. This research was
supported by NSERC of Canada, a Canada Research Chair (L.T.), and
the Canadian Institute for Advanced Research. The work in China was
supported by a grant from the Nature Science
Foundation of China.\\

$^*$ E-mail: louis.taillefer@usherbrooke.ca


\begin{thebibliography}{99}

\bibitem{Lowell} J. Lowell and J. Sousa, J. Low. Temp. Phys. {\bf 3}, 65 (1970).
\bibitem{Sologubenko} A. V. Sologubenko {\it et al.}, Phys. Rev. B {\bf 66}, 014504 (2002).
\bibitem{Boaknin} E. Boaknin {\it et al.}, Phys. Rev. Lett. {\bf 90}, 117003 (2003).
\bibitem{Sutherland} Mike Sutherland {\it et al.}, Phys. Rev. Lett. {\bf 98}, 067003 (2007).
\bibitem{Li} S. Y. Li {\it et al.}, Phys. Rev. Lett. {\bf 99}, 107001 (2007).
\bibitem{Proust} C. Proust {\it et al.}, Phys. Rev. Lett. {\bf 89}, 147003 (2002).
\bibitem{Suzuki} M. Suzuki {\it et al.}, Phys. Rev. Lett. {\bf 88}, 227004 (2002).
\bibitem{Taillefer} L. Taillefer {\it et al.}, Phys. Rev. Lett. {\bf 79}, 483 (1997).
\bibitem{Hawthorn1} D.G. Hawthorn {\it et al.}, Phys. Rev. B {\bf 75}, 104518 (2007).
\bibitem{Sutherland1} Mike Sutherland {\it et al.}, Phys. Rev. B {\bf 67}, 174520 (2003).
\bibitem{Casimir} H. B. G. Casimir, Physica (Amsterdam) {\bf 5}, 495 (1938).
\bibitem{Pohl} R. O. Pohl and B. Stritzker, Phys. Rev. B {\bf 25}, 3608 (1982).
\bibitem{Hurst} W. S. Hurst and D. R. Frankl, Phys. Rev. {\bf 186}, 801 (1969).
\bibitem{Seward} W. D. Seward, Ph.D. thesis, Cornell University, 1965.
\bibitem{Thatcher} P. D. Thatcher, Phys. Rev. {\bf 156}, 975 (1967).
\bibitem{Berman} R. Berman, F. E. Simon, and J. M. Ziman, Proc. R. Soc. London, Ser. A {\bf 220}, 171 (1953).
\bibitem{Sun1} X. F. Sun {\it et al.}, Phys. Rev. B {\bf 72}, 100502(R) (2005).
\bibitem{Sun2} X. F. Sun {\it et al.}, Phys. Rev. Lett. {\bf 96}, 017008 (2006).
\bibitem{Sutherland2} Mike Sutherland {\it et al.}, Phys. Rev. Lett. {\bf 94}, 147004 (2005).
\bibitem{Nicolas} Nicolas Doiron-Leyraud, {\it et al.}, Phys. Rev. Lett. {\bf 97}, 207001 (2006).
\bibitem{Li2} S. Y. Li {\it et al.}, Phys. Rev. Lett. {\bf 95}, 156603 (2005).
\bibitem{Graf} M. J. Graf {\it et al.}, Phys. Rev. B {\bf 53}, 15147 (1996).
\bibitem{Durst} A. C. Durst and P. A. Lee, Phys. Rev. B {\bf 62}, 1270 (2000).
\bibitem{SunandMaki} Y. Sun and K. Maki, Europhys. Lett. {\bf 32}, 355 (1995).
\bibitem{Nakamae} S. Nakamae {\it et al.}, Phys. Rev. B {\bf 63}, 184509 (2001).
\bibitem{Hawthorn} D.G. Hawthorn {\it et al.}, Phys. Rev. Lett. {\bf 90}, 197004 (2003).
\bibitem{Sun3} X.F. Sun {\it et al.}, Phys. Rev. Lett. {\bf 90}, 117004 (2003).


\end{thebibliography}
\end{document}